\documentclass{iau}
\usepackage{graphicx}
\usepackage{natbib}

\usepackage{aas_macros}

\title[The interaction between SMBHs and GCs] 
{The interaction between supermassive black holes and globular clusters}

\author[M. Spera \& M. Arca-Sedda \& R. Capuzzo-Dolcetta]  
{Mario Spera$^{1,2}$ 
 \and Manuel Arca-Sedda$^{2,3}$
 \and Roberto Capuzzo-Dolcetta$^2$ }

\affiliation{$^1$INAF-Osservatorio Astronomico di Padova, Vicolo dell'Osservatorio 5, I-35122, Padova, Italy \\ email: {\tt mario.spera@oapd.inaf.it}  email: {\tt mario.spera@live.it} 
\\ $^2$ Sapienza-Universit\'a di Roma, P.le A. Moro 5, I-00165 Rome, Italy \\
$^3$ Universit\'a di Tor Vergata, Via O. Raimondo 18, I-00173 Rome, Italy}

\pubyear{2015}
\volume{312}  
\pagerange{---}
\jname{---}
\editors{---}
\begin{document}

\maketitle

\newcommand{\msun}{\ensuremath{\,\textrm{M}_{\odot}}}
\newcommand{\lsun}{\ensuremath{\,\textrm{L}_{\odot}}}

\newcommand{\Log}{\ensuremath{\,\textrm{Log}}}

\newcommand{\higpus}{\texttt{HiGPUs}}

\begin{abstract}

Almost all galaxies along the Hubble sequence host a compact massive object (CMO) in their center. The CMO 
can be either a supermassive black hole (SMBH) or a very dense stellar cluster, also known as nuclear star 
cluster (NSC). Generally, heavier galaxies (mass $\gtrsim 10^{11}\msun$) host a central SMBH while lighter 
show a central NSC.  Intermediate mass hosts, instead, contain both a NSC and a SMBH.
One possible formation mechanisms of a NSC relies on the  {\em dry-merger (migratory)} scenario, in which 
globular clusters (GCs) decay toward the center of the host galaxy and merge. In this framework, the absence 
of NSCs in high-mass galaxies can be imputed to destruction of the infalling GCs by the intense tidal field 
of the central SMBH. In this work, we report preliminary results of $N$-body simulations performed using our 
high-resolution, direct, code \higpus, to investigate the effects of a central SMBH on a single GC orbiting 
around it. By varying either the mass of the SMBH and the mass of the host galaxy, we derived an upper limit 
to the mass of the central SMBH, and thus to the mass of the host, above which the formation of a NSC is 
suppressed.

\end{abstract}

\firstsection 
\section{Introduction}
The innermost regions of galaxies often host a very compact stellar cluster with a typical half-light 
radius, $r_{\mathrm{hl}}$, of few parsecs, luminosity $\sim 10^7 \lsun$ and total mass $\sim 10^7 \msun$.  
These compact structures are known as nuclear star clusters (NSCs) \citep{boker2004, cote06, turner2012}. 
NSCs are observed in galaxies of all the Hubble types and, sometimes, they co-exist with a central 
supermassive black hole (SMBH) \citep{graham2009}. For instance, the Milky Way hosts both a NSC with mass 
$\sim 10^7 \msun$ and a SMBH of mass $\sim 4 \times 10^6\msun$ 
\citep{schodel2009}. 

Two (not exclusive) mechanisms have been proposed for NSC formation:

\begin{enumerate}
	\item {\em dissipative}: in this framework the NSC should form thanks to a continuous radial inflow of gas 
and a subsequent, in situ, star formation \citep{milos2004, bekki2007}; 
	\item {\em dissipationless}: this is also known as {\em dry-merging} scenario in which globular clusters 
	(GCs) decay towards the galactic centre via dynamical friction forming and accreting the NSC 
	\citep{tremaine1975, dolcetta1993, dolcetta2008}.
\end{enumerate}

While the in-situ scenario is purely speculative, the dry-merging scenario has been quantitatively 
investigated and matches observational data and correlations \citep{ASCD14b, antonini2012}. Anyway, likely 
the actual formation mechanism can be the result of some combination between the dissipative and the 
dissipationless processes. Moreover, the presence of a central SMBH could significantly alter the process of 
formation and growth of the NSC; many authors have shown that the masses of the SMBH, of the galaxy and of 
the NSC are strongly related. In particular, there is observational evidence of a clear distinction between 
low-mass galaxies (with mass $\lesssim 10^{10}\msun$),whose nuclei are dominated by a NSC,  and heavier 
galaxies, dominated by a SMBH \citep{scott2013}.

In this work, and in the framework of the dry-merging scenario, we investigate the interaction between SMBHs and GCs, in order to determine if the mutual influence may play a role in the co-existence of SMBHs and NSCs. 
We give here preliminary results of a more extended work still in progress \citep{ASCDS15}.

We approach the problem by means of high-precision, direct, $N$-body simulations following the dynamical evolution of several astrophysical systems composed of a galaxy bulge, a central SMBH and a GC moving on different orbits. To this purpose, we used the $N$-body code \higpus 
\citep{dolcetta2013} which, thanks to the hardware acceleration given by graphics processing units, allowed us 
to use $\sim 1\mathrm{M}$ particles in our simulations, obtaining scientific results with high spatial 
resolution.

\section{Model}
\subsection{Globular cluster}
The {\it test} GC in our simulations is built according to a King's mass density profile with central 
potential parameter $W_0=6$, a King's radius $r_{\mathrm{k}}=0.24$ pc and a total mass 
$M_{\mathrm{GC}}=10^6\msun$. 

In all the investigated cases, the GC is initially placed at 50 pc from the central SMBH. Therefore, we 
assume that the GC has already decayed toward the inner galactic region, where the presence of the SMBH may 
play an important dynamical role.
It is worth noting that the 
choice of a large initial GC mass ($10^6\msun$) is motivated by the requirement of orbital shrinking via dynamical friction in less than a Hubble time.
\subsection{Galaxies and central black holes}
Since direct $N$-body simulations cannot handle more than $\sim 10^6$ particles in an efficient way, we decided to model our galaxies by sampling only their central regions. In this work, we focus our attention on elliptical 
galaxies represented as Dehnen's mass density profiles family (at varying $\gamma$, see \cite{dehnen1993}), 
$\rho_{D}(r)$, truncated \citep{mcmillan2007} as

\begin{equation}
\rho_{tr}\left(r\right)=\rho_{D}\left(r\right)\mathrm{sech}\left(\frac{r}{r_{cut}}\right).
\label{eq:trunc}
\end{equation}

The $\rho_{tr}\left(r\right)$ profile falls off as $e^{-r/r_{cut}}$ for $r \gtrsim r_{cut}$ allowing a good representation of the region of interest with a reasonable  number of particles.
The mass value of the central SMBH is assigned according to the formula by \cite{scott2013}
\begin{equation}
\Log\left(\frac{M_{bh}}{\msun}\right)=1.37\Log\left(\frac{M_{gal}}{10^{11}\msun}\right) + 8.06.
\end{equation}
In this work we spanned the mass ranges
$10^{10}\msun<M_{gal}<3.2\times 10^{11} \msun$ and $5\times 10^6 \msun<M_{bh}<5\times 10^8\msun$. Table 
\ref{tab:tab1} summarizes the parameters adopted in each simulation.

\begin{table}
  \begin{center}
  \caption{Parameters of the models. $M_{gal}$ and $M_{gal,cut}$ are the total and truncated (in dependence 
  on $r_{cut}$) galaxy masses, $M_{BH}$ is the SMBH mass; $r_s$ and $\gamma$ is the parameter of the galactic 
  Dehnen's profile; $N_{gal}$ and $N_{GC}$ are the numbers of  particles used in our simulations for the 
  galaxy and for the GC.}
  \label{tab:tab1}
  \begin{tabular*}{\textwidth}{c@{\extracolsep{\fill}}cccccccc}\hline 
 $\mathbf{M_{gal}}$ & $\mathbf{M_{BH}}$ & $\mathbf{r_s}$ & $\mathbf{r_{cut}}$ & $\mathbf{\gamma}$ & 
 $\mathbf{M_{gal,cut}}$ & $\mathbf{N_{gal}}$ & $\mathbf{N_{GC}}$ \\ 
  $\left(\msun\right)$ &  $\left(\msun\right)$ & (kpc) & (pc) & & $\left(\msun\right)$ & & \\ \hline
  
  $10^{10}$            & $5\times 10^6 $ & 0.995 & 70 & 0.3 & $3.4\times 10^7$ & 1,018,742 & 29,832 \\ \hline
  $3.2 \times 10^{10}$ & $2 \times 10^7$ & 1.512 & 70 & 0.3 & $4.1\times 10^7$ & 1,024,025 & 24,550 \\ \hline
  $10^{11}$            & $10^8$          & 1.917 & 70 & 0.2 & $5.9\times 10^7$ & 1,031,338 & 17,237 \\ \hline
  $3.2 \times 10^{11}$ & $5\times 10^8$  & 2.876 & 70 & 0.2 & $6.8\times 10^7$ & 1,033,332 & 15,243 \\ \hline
  \end{tabular*}
 \end{center}
\end{table}

\section{Results for GC circular orbits}
For each galaxy model, we made three simulations corresponding to a circular, an eccentric ($e\sim 0.75$) and a radial orbit of the GC. In this preliminary work we report the results of the GC moving on a circular orbit. 

\begin{figure}
\centering
\includegraphics[scale=0.45]{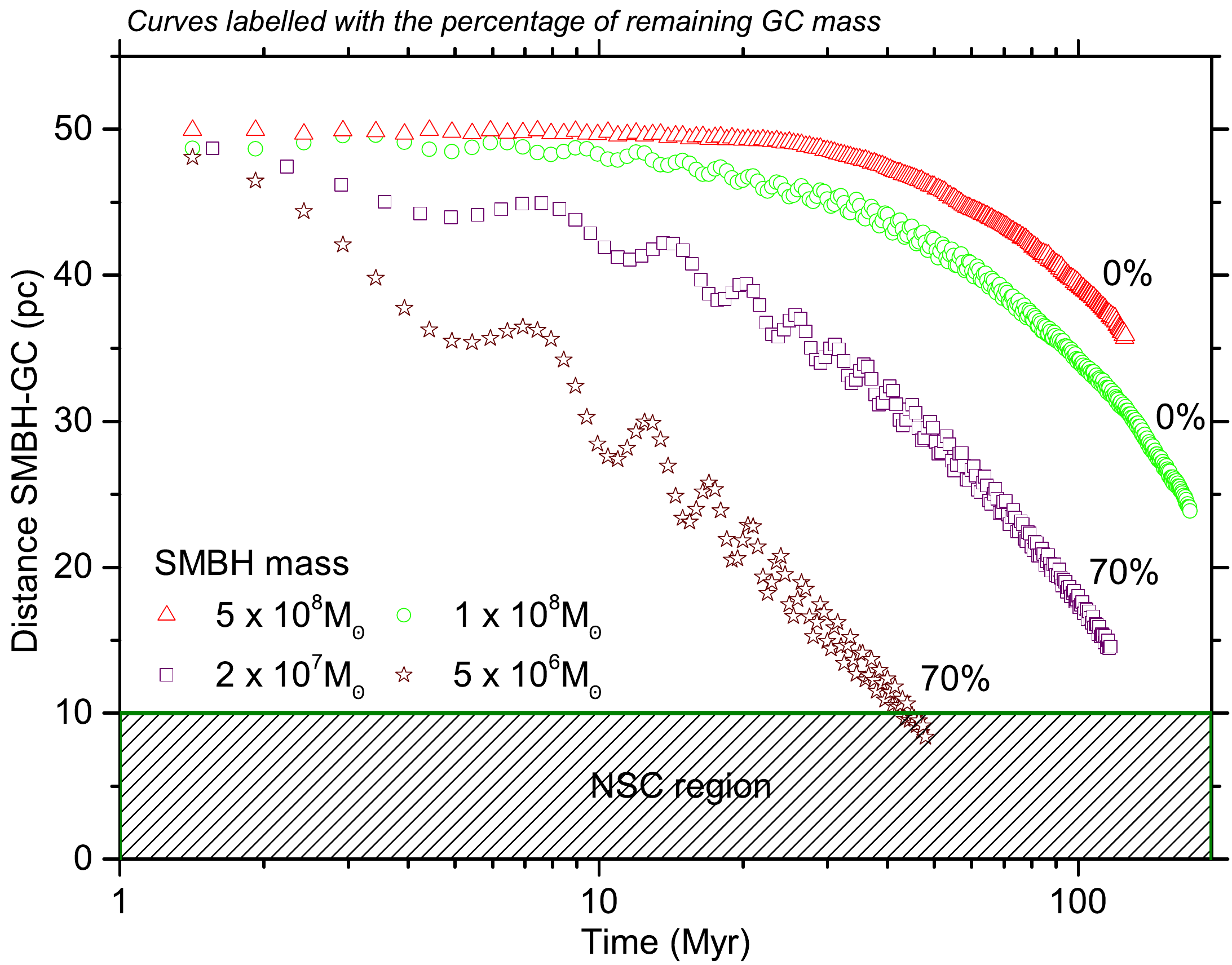}
\caption{Time evolution of the distance of the decaying GC from the SMBH in the models of Table \ref{tab:tab1}. Each curve is labelled with the percentage to the initial mass that the GC keep bound at the end of each simulation.
The shaded area (distances SMBH-GC smaller than 10 pc) marks the region where a NSC may form and grow.}
\label{fig:fig1}
\end{figure}
The four different curves in Fig. \ref{fig:fig1} refer to the time evolution of the GC galactocentric distance in our four models of galaxy. 
Fig. \ref{fig:fig1} shows that heavier SMBHs are able to disrupt
the GC before it gets to the NSC region, while  the incoming GC can survive to the interaction with the central SMBHs if this is relatively light. In fact, in the case of $M_{BH}=5\times 10^6$ M$_\odot$, the GC can come closer than 10 pc to the SMBH preserving 70\% of its initial mass. 

This implies that, in this situation, GCs may indeed 
significantly contribute to the formation and growth of a 
NSC while heavier SMBHs tends to {\em protect} 
the galaxy center, preventing a local mass accumulation and favouring the tidal dissolution of incoming GCs. 
Figure \ref{fig:fig1} provides also evidence of a transition regime between {\em dynamical friction 
dominated} 
galaxies (formation and growth of NSC by mergers) and {\em tidal disruption dominated} galaxies (no NSC). Our, preliminary, simulations suggest the transition lies in the range of SMBHs masses between $2 \times 10^7 \msun$ 
and $10^8 \msun$.

\section{Conclusions}
We presented preliminary results about the problem of the dearth of NSCs in high mass (elliptical) galaxies.
We performed some high-precision, direct $N$-body simulations to investigate the dynamical fate of a massive 
GC orbiting the inner region of a galaxy.  
We showed that a SMBH heavier than $\sim 10^8 \msun$ can efficiently disrupt the infalling GC before it gets 
to what we called {\em NSC region}, and therefore it may inhibit the formation process, by subsequent merging 
events, of a NSC.
On the other hand, the incoming GC survives the interaction with lighter BHs and, thus, can contribute to the formation and growth of a NSC. 
To conclude, our simulation results are a reliable confirmation of the important role played by a massive central black hole on the infalling GCs, as first pointed out by \citet{dolcetta1993} and, more recently, by \citet{antonini2013}.
Nevertheless, a firmer statement about the topic studied here deserves:
\begin{enumerate}
\item a wider range of initial conditions for both the GC structure and its initial orbital parameters;
\item an extension of the galaxy models, to determine more precisely the threshold in $M_{BH}$ below which 
the dry-merging of GCs easily allows the formation of a NSC;
\item a better $N$-body sampling of our models, possible by taking advantage of the next generation hardware and software.
\end{enumerate} 

\section{Acknowledgements}
MS thanks M. Mapelli for useful discussions, and acknowledges financial support from the MIUR through grant 
FIRB 2012 RBFR12PM1F. MAS acknowledges financial support provided by MIUR through the grant PRIN 2010 LY5N2T 
005.


\begin{thebibliography}{}

\bibitem[Antonini(2013)]{antonini2013} Antonini, F. 2013, \apj, 763, 62

\bibitem[Antonini et al.(2012)]{antonini2012} Antonini, F., Capuzzo-Dolcetta, R., Mastrobuono-Battisti, A., \& Merritt, D.\ 2012, \apj, 750, 111

\bibitem[Arca-Sedda \& Capuzzo-Dolcetta (2014)]{ASCD14b} Arca-Sedda, M., \& Capuzzo-Dolcetta, R. \ 2014, \mnras, 444, 3738-3755

\bibitem[Arca-Sedda et al.(2015)]{ASCDS15} Arca-Sedda, M. , Capuzzo-Dolcetta, R., \& Spera, M. 2015, in preparation

\bibitem[Bekki(2007)]{bekki2007} Bekki, K.\ 2007, \pasa, 24, 77
 
\bibitem[B{\"o}ker et al.(2004)]{boker2004} B{\"o}ker, T., Sarzi, M., McLaughlin, D.~E., et al.\ 2004, \aj, 127, 105

\bibitem[Capuzzo-Dolcetta (1993)]{dolcetta1993} Capuzzo-Dolcetta, R.\ 1993, \apj, 415, 616

\bibitem[Capuzzo-Dolcetta \& Miocchi (2008)]{dolcetta2008} Capuzzo-Dolcetta, R., \& Miocchi, P.\ 2008, \apj, 681, 1136

\bibitem[Capuzzo-Dolcetta et al.(2013)]{dolcetta2013} Capuzzo-Dolcetta, R., Spera, M., \& Punzo, D.\ 2013, JCP, 236, 580

\bibitem[C{\^o}t{\'e} et al.(2006)]{cote06} {C{\^o}t{\'e}}, P., {Piatek}, S., {Ferrarese}, L., et al.\ 2006, \apjs, 165, 57-94

\bibitem[Dehnen(1993)]{dehnen1993} Dehnen, W.\ 1993, \mnras, 265, 250 

\bibitem[Graham \& Spitler(2009)]{graham2009} Graham, A.~W., \& Spitler, L.~R.\ 2009, \mnras, 397, 2148

\bibitem[McMillan \& Dehnen(2007)]{mcmillan2007} McMillan, P.~J., \& Dehnen, W.\ 2007, \mnras, 378, 541

\bibitem[Milosavljevi{\'c}(2004)]{milos2004} Milosavljevi{\'c}, M.\ 2004, \apjl, 605, L13

\bibitem[Sch{\"o}del et al.(2009)]{schodel2009} Sch{\"o}del, R., Merritt, D., \& Eckart, A.\ 2009, \aap, 502, 91

\bibitem[Scott \& Graham(2013)]{scott2013} Scott, N., \& Graham, A.~W.\ 2013, \apj, 763, 76

\bibitem[Tremaine et al.(1975)]{tremaine1975} Tremaine, S.~D., Ostriker, J.~P., \& Spitzer, L., Jr.\ 1975, \apj, 196, 407

\bibitem[Turner et al.(2012)]{turner2012} Turner, M.~L., C{\^o}t{\'e}, P., Ferrarese, L., et al.\ 2012, \apjs, 203, 5

\end{thebibliography}
\end{document}